# A Generalized Deletion Machine

[1] Indranil Chakrabarty, [2] Satyabrata Adhikari

[1] Department of Mathematics, Heritage Institute of Technology, Chowbhaga Road, Anandapur, Kolkata-107

[2] Department of Mathematics, Bengal Engineering and Science University, Sibpur, Howrah-711103

**Abstract:** In this work we prescribe a more generalized quantum-deleting machine (input state dependent). The fidelity of deletion is dependent on some machine parameters such that on alteration of machine parameters we get back to standard deleting machines. We also carried out a various comparative study of various kinds of quantum deleting machines. We also plotted graphs, making a comparative study of fidelity of deletion of the deletion machines, obtained as particular cases on changing the machine parameters of our machine.

**INTRODUCTION:** In the arena of quantum information theory it is well known fact that any arbitrary quantum state can't be copied or deleted perfectly [1,2]. Linearity of quantum mechanics doesn't allow perfect replication and deletion of an arbitrary quantum state. But this doesn't rules out the possibility of designing approximate cloning and deletion machine [3,8,9,10]. Cloning machine can be divided into two general categories.(1)State dependent cloning machine: a cloning machine that depends on input state such as Wooters –Zurek cloning machine. (2)Universal Quantum cloning machine: a cloning machine which is independent of input state such as Buzek-Hillary cloning machine. Not only that quantum state can exactly be copied by unitary and measurement process it had also been done that non-orthogonal quantum state can exactly be copied by unitary and measurement process [4]. It had also been done that non-orthogonal states can be obtained as a linear superposition of multiple copies of novel cloning machine [5].In the age of information there had been lot of speculation regarding the possibility and impossibility of certain quantum operation on an unknown quantum state [6]. . The most striking feature is that the process of deletion is not at all same as the erasure. When we want to get rid of last bit of information, we can do it by spending certain amount of energy known as Landuer's erasure principle [7]. But on the contrary deleting of a quantum state is more or less uncopying type of operation which basically starts up with two identical quantum state and as a result of deletion one of the state gets swapped with some ancilla state. Many works had been carried out in the field of quantum deletion. No Deleting principle had been generalized for higher dimensional for higher dimensional quantum state by Pati and Braunstein [8].A non-optimal quantum-deleting machine was proposed by D.Qiu [9].The striking feature of this deletion machine is that .the fidelity of deletion in the ancilla mode is independent of the input state parameters. Adhikary and Chowdhury proposed a Deletion machine (input state dependent) giving a much better fidelity of deletion and minimum input state distortion [10].Not only that authors along with the introduction of new kind of deletion machine also applied their deletion machine on imperfect cloned copies and made comparative study with standard deletion machines.

In this work we construct a more generalized deleting machine, which is input state dependent. We will then apply the prescribed deletion operation on an unknown quantum state and due to this operation we find that the input state is distorted, so we will calculate the amount of distortion in the original mode and the fidelity of deletion for the deleted mode and we see that both of it depends not only on the input states but also on the machine parameters. Now by suitable choice of the machine parameters we get a deletion machine with high fidelity of deletion and minimum average distortion of the input state.

## 2. INPUT STATE DEPENDENT DELETION MACHINE:

In this work we prescribe a deleting machine given by,

$$U|0\rangle|0\rangle|Q\rangle \to |0\rangle|\Sigma\rangle|A_0\rangle$$
$$U|0\rangle|1\rangle|Q\rangle \to (a_0|0\rangle|1\rangle + b_0|1\rangle|0\rangle + c_0|0\rangle|0\rangle + d_0|1\rangle|1\rangle)|Q\rangle$$
$$U|1\rangle|0\rangle|Q\rangle \to (a_1|0\rangle|1\rangle + b_1|1\rangle|0\rangle + c_1|0\rangle|0\rangle + d_1|1\rangle|1\rangle)|Q\rangle$$
$$U|1\rangle|1\rangle|Q\rangle \to |1\rangle|\Sigma\rangle|A_1\rangle$$

(2.1)

(where $\{|Q\rangle, |A_0\rangle, |A_1\rangle, |\Sigma\rangle\}$ have their usual meaning and $a_i, b_i, c_i, d_i; i = \{0,1\}$ are complex numbers)
Due to unitarity of the transformation the following conditions will hold,

$$\langle A_i | A_i \rangle = 1$$
$$|a_i|^2 + |b_i|^2 + |c_i|^2 + |d_i|^2 = 1$$
$$a_i a^*_{1-i} + b_i b^*_{1-i} + c_i c^*_{1-i} + d_i d^*_{1-i} = 0$$
$$\langle A_1 | Q \rangle = \langle A_0 | Q \rangle = 0$$

(2.2)

Further we assume that,
$$\langle A_1 | A_0 \rangle = \langle A_0 | A_1 \rangle = 0$$ (2.3)

A general pure state is given by,
$$|\Psi\rangle = \alpha|0\rangle + \beta|1\rangle; (\alpha^2 + \beta^2 = 1)$$ (2.4)

(without any loss of generality we can assume that $\alpha, \beta$ are real numbers)
The density operator described the pure state (2.4) is given by
$$\rho_1^{ID} = \alpha^2|0\rangle\langle 0| + \alpha\beta|0\rangle\langle 1| + \alpha\beta|1\rangle\langle 0| + \beta^2|1\rangle\langle 1|$$ (2.5)

The output state after operating deletion machine is given by
$$|\Psi\rangle_{123}^{OUT} \equiv U|\Psi\rangle|\Psi\rangle|Q\rangle = \alpha^2|0\rangle|\Sigma\rangle|A_0\rangle + \alpha\beta[g|0\rangle|1\rangle + h|1\rangle|0\rangle + e|0\rangle|0\rangle + f|1\rangle|1\rangle]|Q\rangle + \beta^2|1\rangle|\Sigma\rangle|A_1\rangle$$ (2.6)
where, $g = a_0 + a_1, h = b_0 + b_1, e = c_0 + c_1, f = d_0 + d_1$ (2.7)

The reduced density operator of the output state in mode 1 is given by,
$$\rho_1^{OUT} = Tr_{23}(|\Psi\rangle_{123}^{OUT}\,_{123}^{OUT}\langle\Psi|) = [\alpha^4 + \alpha^2\beta^2 gg^* + \alpha^2\beta^2 ee^*]|0\rangle\langle 0| + [\alpha^2\beta^2 fg^* + \alpha^2\beta^2 e^*h]|1\rangle\langle 0|$$
$$+ [\alpha^2\beta^2 eh^* + \alpha^2\beta^2 gf^*]|0\rangle\langle 1| + [\beta^4 + \alpha^2\beta^2 ff^* + \alpha^2\beta^2 hh^*]|1\rangle\langle 1|$$

(2.8)

The reduced density operator of the output state in mode 2 is given by,
$$\rho_2^{OUT} = \alpha^4|\Sigma\rangle\langle\Sigma| + gg^*\alpha^2\beta^2|1\rangle\langle 1| + g^*e\alpha^2\beta^2|0\rangle\langle 1| + hh^*\alpha^2\beta^2|0\rangle\langle 0| + h^*f\alpha^2\beta^2|1\rangle\langle 0|$$
$$+ e^*g\alpha^2\beta^2|1\rangle\langle 0| + ee^*\alpha^2\beta^2|0\rangle\langle 0| + hf^*\alpha^2\beta^2|0\rangle\langle 1| + ff^*\alpha^2\beta^2|1\rangle\langle 1| + \beta^4|\Sigma\rangle\langle\Sigma|$$

(2.9)

Now to see the performance of the prescribed machine we must calculate the distortion of the input state and fidelity of deletion. Therefore the distance between the density operators $\rho_1^{ID}, \rho_1^{OUT}$ is given by,

$$D_1(\alpha^2) = Tr[\rho_1^{ID} - \rho_1^{OUT}]^2$$
$$= L(\alpha^4\beta^4) - 2(M_3 + M_4)\alpha^3\beta^3 + 2\alpha^2\beta^2$$

(2.10)

where, $L = K + 2M_3 M_4$ (2.11)

$$K = (M_1 - 1)^2 + (M_2 - 1)^2 \tag{2.12}$$

$$\begin{aligned} M_1 &= ee^* + gg^* \\ M_2 &= hh^* + ff^* \\ M_3 &= eh^* + gf^* \\ M_4 &= he^* + fg^* \end{aligned} \tag{2.13}$$

Since, $D_1$ depends on $\alpha^2$, so average distortion of the input qubit in the mode 1 is given by

$$\overline{D_1} = \int_0^1 D_1(\alpha^2) d\alpha^2$$
$$= \frac{L}{30} + \frac{1}{3} - (0.589)(M_3 + M_4) \tag{2.14}$$

The reduced density matrix of the qubit in the mode 2 contains error due to imperfect deleting and the error can be measured by calculating the fidelity.
Thus the fidelity is given by,

$$F_1 = \langle \Sigma | \rho_2^{OUT} | \Sigma \rangle = 1 - K_1 \alpha^2 \beta^2 \tag{2.15}$$

where $K_1 = 2 - \{[gg^* + ff^*](M_1')^2 + [hh^* + ee^*][1 - (M_1')^2] + (M_1')(\sqrt{1 - (M_1')^2})[g^*e + h^*f + e^*g + hf^*]\}$ (2.16)

$$M_1' = \langle \Sigma | 0 \rangle \tag{2.17}$$

Since, the fidelity of deletion depends on the input state, so the average fidelity over all the input state is given by,

$$\overline{F_1} = \int_0^1 F_1(\alpha^2) d\alpha^2 = 1 - \frac{K_1}{6}; (0 < K_1 < 6) \tag{2.18}$$

Now let us examine four possible cases:

**CASE 1:**
If $e = f = g = h = 0$, then $K_1 = 2, L = 2, M_3 = M_4 = 0$. (2.19)
Therefore, the average distortion and the average fidelity of deletion are given by,

$$\overline{D_1} = \frac{2}{5}, \overline{F_1} = \frac{2}{3} \tag{2.20}$$

**CASE 2:**
If $g = 0, h = 0$ and $|f|^2 = 1, |e|^2 = 1$, then $K_1 = 1, L = 0, M_3 = M_4 = 0$ (2.21)
and assume, , then. Therefore average distortion and the average fidelity of deletion are given by,

$$\overline{D_1} = \frac{1}{3}, \overline{F_1} = \frac{5}{6} \tag{2.22}$$

Now from the above equation it is very clear that if we choose the machine parameters in such a way so that we are able to keep the distortion in minimum level and increase the average fidelity to $5/6$.

**CASE 3:**
If $a_0 = 1, a_1 = 0, b_0 = 0, b_1 = 1, c_0 = c_1 = d_0 = d_1 = 0$ (2.23)
i.e. $g = h = 1, e = f = 0$, then $K_1 = 1, L = 0, M_3 = M_4 = 0$.
The average distortion and average fidelity are the same as in case 2 but the above selection of the machine parameters reduces the general deletion machine to Pati-Braunstein deletion machine [2,8].

**CASE 4**:
If $c_0 = c_1 = d_0 = d_1 = 0$, then the general deletion machine reduces to Adhikari et.al. [10] deletion machine where the average distortion and average fidelity are given by

$$\overline{D_1} = \frac{N}{30} + \frac{1}{3}; N = (gg^* - 1)^2 + (hh^* - 1)^2$$
$$\overline{F_1} = 1 - \frac{K_2}{6}; K_2 = 2 - \{gg^*(M_1')^2 + hh^*[1 - (M_1')^2]\}; M_1' = \langle \Sigma | 0 \rangle$$
(2.24)

Further if we choose the machine parameters g and h in such a way that $|g|^2 = 1$ and $|h|^2 = 1$, then the average distortion and the average fidelity of deletion are given by,
$$\overline{D_1} = 1/3, \overline{F_1} = 5/6.$$ (2.25)

Thus we have constructed a generalized deleting machine and find that it's fidelity of deletion and input state distortion not only depends on input state but also on the machine parameters. Thus we had also shown that for certain values for machine parameters, the deleting machine is converted to standard deleting machines like P-B deleting machine etc.

**FIDELITY OF DELETION**

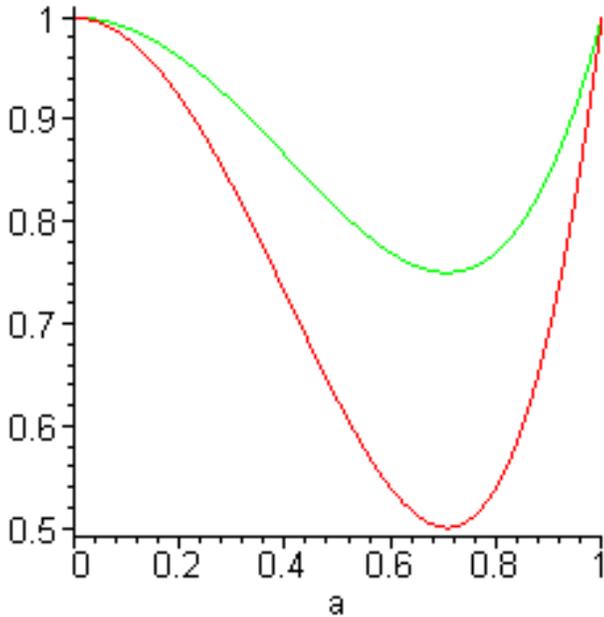

In the above graph the fidelity of deletion is plotted along Y-axis and $\alpha^2$ along X-axis. Green line gives the variation of fidelity of deletion $F_1$ with $\alpha^2$ for the last three cases while the red line indicates the variation of fidelity of deletion $F_1$ with $\alpha^2$ for the first case.

### 3. CONCLUSION:

In this work our objective is to construct a generalized deletion machine, such that standard deletion Machines like P-B deletion machine, etc may be obtained as special cases of our deletion machine. Though the machine proposed by us doesn't give low distortion and high fidelity than the deletion machines proposed earlier, but on alteration of the machine parameters it reduces to standard deletion machines like P-B deleting machines etc.

### 4. ACKNOWLEDGEMENT:

Authors acknowledge Prof.B.S.Chowdhury for his encouragement and valuable discussions in completing this work. The second author wishes to acknowledge CSIR (project number 8/3(38)/2003-EMR-1) for providing financial support in completion of this work.